\documentclass[runningheads]{llncs}
\usepackage{graphicx}

\usepackage{multirow}

\usepackage{cite}

\usepackage{amssymb}

\begin{document}
\title{Dual-Task Mutual Learning for Semi-Supervised Medical Image Segmentation} 
\titlerunning{Dual-Task Mutual Learning for Semi-Supervised Segmentation}
%

\author{Yichi Zhang\inst{1} \and Jicong Zhang\inst{1,2,3,4}}

\authorrunning{Y. Zhang \textit{et al.}}
%
\institute{School of Biological Science and Medical Engineering, Beihang University, Beijing, China \and Hefei Innovation Research Institute, Beihang University, Hefei, China \and Beijing Advanced Innovation Centre for Biomedical Engineering, Beijing, China \and Beijing Advanced Innovation Centre for Big Data-Based Precision Medicine, Beijing, China}

\maketitle              

\begin{abstract}
The success of deep learning methods in medical image segmentation tasks usually requires a large amount of labeled data. However, obtaining reliable annotations is expensive and time-consuming. Semi-supervised learning has attracted much attention in medical image segmentation by taking the advantage of unlabeled data which is much easier to acquire. 
In this paper, we propose a novel dual-task mutual learning framework for semi-supervised medical image segmentation. 
Our framework can be formulated as an integration of two individual segmentation networks based on two tasks: learning region-based shape constraint and learning boundary-based surface mismatch.
Different from the one-way transfer between teacher and student networks, an ensemble of dual-task students can learn collaboratively and implicitly explore useful knowledge from each other during the training process.
By jointly learning the segmentation probability maps and signed distance maps of targets, our framework can enforce the geometric shape constraint and learn more reliable information. 
Experimental results demonstrate that our method achieves performance gains by leveraging unlabeled data and outperforms the state-of-the-art semi-supervised segmentation methods.
\renewcommand{\thefootnote}{}\footnote{Our code is available at https://github.com/YichiZhang98/DTML}

\end{abstract}
\section{Introduction}

Medical image segmentation aims to understand images in pixel-level and label each pixel into a certain class, which is a fundamental step for many clinical applications \cite{van2011computer,sykes2014reflections}.
Recently, deep learning techniques have showed significant improvements and achieved state-of-the-art performances in many medical image segmentation tasks \cite{bernard2018deep,ma2020abdomenct,heller2020state}.
However, training deep neural networks usually relies on massive labeled dataset, while it is extremely expensive and time-consuming to obtain large-amount of well-annotated data where only professional experts can provide reliable annotations for medical imaging.
To reduce the labeling cost, many studies have focused on developing annotation-efficient medical image segmentation methods with scarce annotations or weak annotations \cite{tajbakhsh2020embracing,cheplygina2019not,zhang2020exploiting,wang2020annotation}.
For medical imaging, unlabeled data is much easier to acquire and can be used in conjunction with labeled data to train segmentation models.
As a result, semi-supervised learning has been widely explored to learn from a limited amount of labeled data and an arbitrary amount of unlabeled data \cite{qi2020small,li2020transformation}, which is a fundamental, challenging problem and has a high impact on real-world clinical applications.

In this paper, we propose a novel dual-task mutual learning framework for semi-supervised medical image segmentation. 
The framework can be formulated as an integration of two individual segmentation networks based on different tasks.
The segmentation task aims at generating segmentation probabilistic maps while the regression task aims at regressing the signed distance maps.
Since the output of different tasks can be mapped to the same predefined space, dual-task networks can learn different representations of segmentation targets from different perspectives. 
Following the mutual learning manner \cite{zhang2018deep}, we aim at encouraging dual-task networks to learn collaboratively and explore useful knowledge from each other during the training process.
Each network is primarily directed by a conventional supervised learning loss for training. With supervised learning, both networks learn reliable representation of the segmentation task from different task-level condition, therefore estimate the probabilities of the most likely categories differently. 
Under the semi-supervised learning setting, we activate the unsupervised cross-task consistency loss to encourage consistent predictions of the same input in order to utilize the unlabeled data. By jointly learning the segmentation probability maps and signed distance maps, our framework can enforce the geometric shape constraint and collaboratively learn more reliable information for segmentation throughout the training procedure. 

Our method is evaluated on the Atrial Segmentation Challenge dataset for left atrium (LA) segmentation \cite{xiong2021global} with extensive comparisons to existing methods. The experimental results demonstrate that our method achieves performance gains by leveraging unlabeled data and outperforms the state-of-the-art semi-supervised segmentation methods.

\section{Related Work}

\subsection{Semi-Supervised Medical Image Segmentation}

To utilize unlabeled data for semi-supervised medical image segmentation, a simple and intuitive method is to assign pseudo annotations to unlabeled data and then train the segmentation model using both labeled and pseudo labeled data. Pseudo annotations are commonly generated in an iterative approach wherein a model iteratively improves the quality of pseudo annotations by learning from its own predictions on unlabeled data \cite{bai2017semi}. Zhang \textit{et al.} \cite{zhang2017deep} introduced adversarial learning for biomedical image segmentation by encouraging the segmentation output of unlabeled data to be similar to annotations of labeled data. Although semi-supervised learning with pseudo annotations has shown promising performance, model-generated annotations can still be noisy and has detrimental effects to the subsequent segmentation model \cite{min2019two,nie2018asdnet}. 

Recent efforts in semi-supervised segmentation have been focused on incorporating unlabeled data into the training procedure with an unsupervised loss function. 
Some of these methods enforce the consistency between model predictions on the original data and the perturbed data by adding small perturbations to the unlabeled data. For example, Li \textit{et al.} \cite{li2020transformation} proposed to constrain the consistency under transformation like rotation to utilize unlabeled data. Yu \textit{et al.} \cite{yu2019uncertainty} extended the mean teacher paradigm \cite{tarvainen2017mean} with the guidance of uncertainty map for semi-supervised learning. \cite{fang2020dmnet} proposed to utilize unlabeled data by minimizing the difference between soft masks generated by two decoders. Some other methods focus on enforcing the similar distribution of predictions using an adversarial loss \cite{nie2018asdnet,li2020shape}.

\subsection{Signed Distance Maps}

Different from most existing segmentation methods that use binary or multi-label mask as ground truth, signed distance maps (SDM) can provide an alternative to classical ground truth by transforming binary masks to gray-level images where the intensities of pixels are changed according to the distance to the closest boundary \cite{ma2020distance}, which has been applied for medical image segmentation tasks to obtain further improvements by offering an implicit representation of the ground truth \cite{navarro2019shape,dangi2019distance}. 
A commonly used method is adding auxiliary regression head to the end of classic encoder-decoder network to generate signed distance maps. 
Specifically, the network can be divided into two branches to generate the segmentation probabilistic maps and regress the signed distance maps at the same time. Due to the task difference, these two branches can learn different representations of segmentation targets from different perspectives, so as to obtain further improvements.
On the condition that the output of different tasks can be mapped to the same predefined space, instead of existing data-level regularization methods, we focus on building task-level regularization to utilize unlabeled data in our framework.

\begin{figure}[t]
	\includegraphics[width=12.5cm]{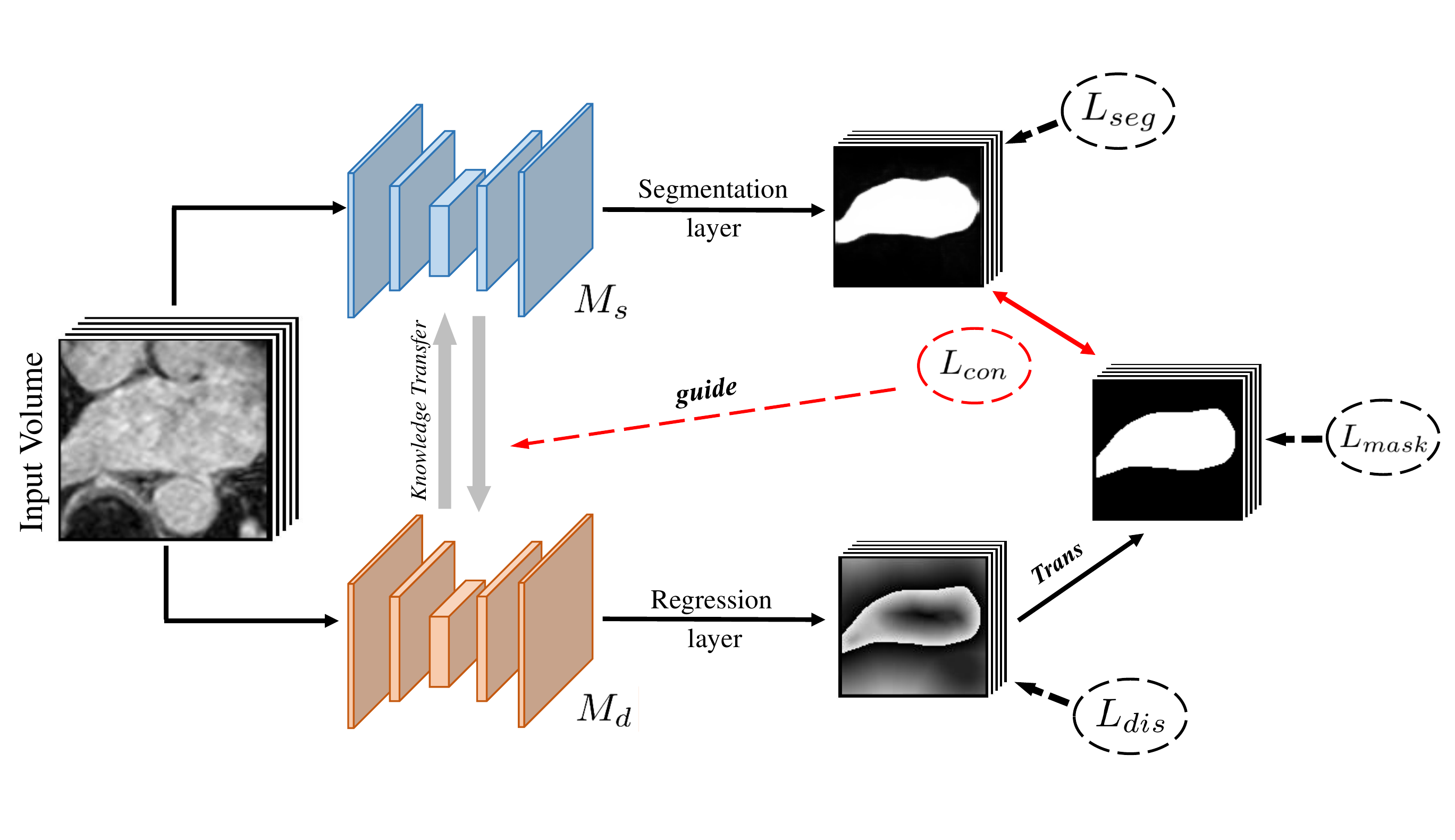}
	\caption{The overview of our proposed dual-task mutual learning framework for semi-supervised medical image segmentation. The framework can be formulated as an integration of two individual segmentation networks, the upper one named $M_{s}$ for generating the segmentation probabilistic maps and the lower one named $M_{d}$ for regressing the signed distance maps. These two networks share the same backbone structure, and have task specific segmentation/regression output layers. During the training process, the networks are optimized with mutual learning manner to implicitly explore useful knowledge from each other.}
	\label{Fig1}
\end{figure}

\section{Method}

In this section, we introduce details about our proposed dual-task mutual learning framework.
As illustrated in Figure \ref{Fig1}, our framework can be formulated as an integration of two individual segmentation networks. These two networks share the same backbone structure and have specific segmentation and regression heads for different tasks.
The upper one $M_{s}$ aims at generating segmentation probabilistic maps while the lower one $M_{d}$ aims at regressing the signed distance maps.
Since the output of different tasks can be mapped to the same predefined space, dual-task networks can learn different representations of segmentation targets from different perspectives. We focus on building task-level regularization to enable each network learn from peer network's guidance and introduce a cross-task consistency regularization to enforce the representation of predictions to be consistent. To fully utilize the spatial information, we task 3D volumes as the input for both networks.

\subsection{Dual-Task Networks}

Following the design of encoder-decoder architecture \cite{ronneberger2015u,cciccek20163d,milletari2016v} to generate segmentation probabilistic maps, an auxiliary regression head is added to generate the signed distance maps composed by a 3D convolution block followed by the \textit{tanh} activation. The signed distance maps of ground truth $G$ can be defined by

\begin{equation}
G_{SDF}=\left\{
\begin{array}{lllll}
-\inf \limits_{y \in \partial G}\|x-y\|_{2} , & x \in G_{\mathrm{in}}  \\
\\
 0, & x \in \partial G \\
 \\
+\inf \limits_{y \in \partial G}\|x-y\|_{2} , & x \in G_{\mathrm{out}}  \\
\end{array}\right.
\end{equation}
where $\|x-y\|$ is the Euclidian distance between voxels $x$ and $y$, and $G_{in}$, $\partial G$, $G_{out}$ represents the inside, boundary and outside of the target object. In general, $G_{SDF}$ takes negative values inside the target and positive values outside the object, and the absolute value is defined by the distance to the closest boundary point.
To transform the output of signed distance maps to segmentation output, we utilize a smooth approximation to the inverse transform as in \cite{luo2020semi}, which can be defined by

\begin{equation} \label{trans}
G_{mask}=\frac{1}{1+e^{-k \cdot z}} , z \in G_{SDF}
\end{equation}
where $z$ is the value of signed distance maps at voxel $x$, and $k$ is a transform factor selected as large as possible to approximate the transform.
Our dual-task networks share the same backbone structure and differ at the end of the network. Specifically, for $M_{s}$ only the segmentation head is activated, while for $M_{d}$ only the regression head is activated.
Given an input image $X \in R^{H \times W \times D}$, dual-task networks $M_{s}$ and $M_{d}$ generate the confidence score map $\hat Y_{seg} \in [0,1]^{H \times W \times D}$ and signed distance map $\hat Y_{dis} \in \textbf{R}^{H \times W \times D}$ as follows

\begin{equation}
\hat Y_{seg} = f_{seg}(X;\theta_{seg}) , \quad \hat Y_{dis} = f_{dis}(X;\theta_{dis})
\end{equation}
where $\theta_{seg}$, $\theta_{dis}$ are corresponding parameters of segmentation network $M_{s}$ and regression network $M_{d}$, respectively.

\subsection{Mutual Learning for Semi-Supervised Segmentation}

For semi-supervised segmentation of 3D medical images, where the training set $\mathcal{D}$ contains $M$ labeled cases and $N$ unlabeled cases, we denote the labeled set as $\mathcal{D}_{L} = \{X_{i}, Y_{i}\}_{i=1}^{M}$ and the unlabeled set as $\mathcal{D}_{U} = \{X_{i}\}_{i=1}^{N}$, where $X_{i} \in \textbf{R}^{H \times W \times D}$ is the input volume and $Y_{i} \in \{0,1\}^{H \times W \times D}$ is the corresponding ground truth. 

For labeled cases, each network is primarily directed by supervised loss to learn reliable representation of the segmentation task.
We employ the combination of dice loss and cross-entropy loss as the supervised loss $\mathcal{L}_{seg}$ for the segmentation of $M_{s}$.
While for $M_{d}$, two options of supervision can be applied for the training.
The first choice is using $\mathcal{L}_{2}$ loss between the output signed distance maps and transformed distance maps of ground truth named $\mathcal{L}_{dis}$ as the supervision. Another choice is using dice loss between transformed segmentation mask and the ground truth named $\mathcal{L}_{mask}$. 
We empirically found that supervised directly on segmentation masks with $\mathcal{L}_{mask}$ can achieve better performance.
Under the semi-supervised learning setting, to utilize unlabeled cases for training, we activate the unsupervised cross-task consistency loss to encourage consistent predictions of the same input in order to utilize the unlabeled data. By jointly learning the segmentation probability maps and signed distance maps, our framework can enforce the geometric shape constraint and collaboratively learn more reliable information for segmentation throughout the training procedure. 
To quantify the match of the two network's predictions, we focus on enforcing cross-task consistency between segmentation predictions and transformed regression predictions: 

\begin{equation}
\mathcal{L}_{con} = \lambda_{con} \| f_{seg}(X;\theta_{seg}) - f^{-1}_{mask}( f_{dis}(X;\theta_{dis})) \| ^{2}
\end{equation}
where $f^{-1}_{mask}$ is the transformation of signed distance maps to segmentation maps as described in (\ref{trans}), and $\lambda_{con}$ is the ramp-up weighting coefficient to control the trade-off between the segmentation loss and consistency loss. Following \cite{yu2019uncertainty}, we use a Gaussian ramp-up function $ \lambda_{con}(t)=0.1*e^{-5(1-T/T_{max})} $ in all our experiments where t represents the number of iterations.

Therefore, the goal of our semi-supervised segmentation framework is to minimize the following combined functions.
\begin{equation}
\min \limits_{\theta_{seg}} \sum_{i\in \mathcal{D}_{L}} \mathcal{L}_{seg}(f_{seg}(X_{i};\theta_{seg}),Y_{i}) + 
\sum_{i\in \mathcal{D}} \mathcal{L}_{con}(f^{-1}_{mask}(f_{dis}(X_{i};\theta_{dis})),f_{seg}(X_{i};\theta_{seg}))
\end{equation}
\begin{equation}
\min \limits_{\theta_{dis}} \sum_{i\in \mathcal{D}_{L}} \mathcal{L}_{mask}(f^{-1}_{mask}(f_{dis}(X_{i};\theta_{dis})),Y_{i}) + 
\sum_{i\in \mathcal{D}} \mathcal{L}_{con}(f^{-1}_{mask}(f_{dis}(X_{i};\theta_{dis})),f_{seg}(X_{i};\theta_{seg}))
\end{equation}

\section{Experiments}

\subsection{Dataset and Implementation Details}

We evaluate our method on the Left Atrium (LA) dataset from Atrial Segmentation Challenge\renewcommand{\thefootnote}{1}\footnote{http://atriaseg2018.cardiacatlas.org/data/} \cite{xiong2021global}. 
The dataset contains 100 3D gadolinium-enhanced MR imaging scans (GE-MRIs) and corresponding LA segmentation mask for training and validation. These scans have an isotropic resolution of $0.625 \times 0.625 \times 0.625$ $mm^{3}$. Following the task setting in \cite{yu2019uncertainty}, we split the 100 scans into 80 scans for training and 20 scans for testing, and apply the same pre-processing methods. Out of the 80 training scans, we use the same 20\%/16 scans as labeled data and the remaining 80\%/64 scans as unlabeled data for semi-supervised segmentation task. 

Our framework is implemented in PyTorch, using an NVIDIA Tesla V100 GPU.
In this work, we use V-Net \cite{milletari2016v} as the backbone structure for all experiments to ensure a fair comparison. To incorporate signed distance maps for mutual learning, an auxiliary regression head is added at the end of the original V-Net.
We use the Stochastic Gradient Descent (SGD) optimizer to update the network parameters for 6000 iterations, with an initial learning rate (lr) 0.01 decayed
by 0.1 every 2500 iterations. The batch size is 4, consisting of 2 labeled images and 2 unlabeled images. We randomly crop $112 \times 112 \times 80$ sub-volumes as the network input and the final segmentation results are obtained using a sliding window strategy.
We use the standard data augmentation techniques on-the-fly to avoid overfitting during the training procedure \cite{yu2017automatic}, including randomly flipping, and rotating with 90, 180 and 270 degrees along the axial plane.

We use four complementary evaluation metrics to quantitatively evaluate the segmentation results. Dice similarity coefficient (Dice) and Jaccard Index (Jaccard), two region-based metrics, are used to measure the region mismatch. Average surface distance (ASD) and 95\% Hausdorff Distance (95HD), two boundary-based metrics, are used to evaluate the boundary errors between the segmentation results and the ground truth.

\begin{table}[t]
	\caption{Comparison of different supervised loss functions for our dual-task mutual learning framework. All the models follow the same task setting with 16 labeled scans and 64 unlabeled scans for training. } 	\label{Table1}
	\centering
	\renewcommand\arraystretch{1.1}
	\begin{tabular}{c|c|c|c|c}
		\hline 	\hline
		\multirow{2}{*}{\bf{Supervised Loss $M_{s}/M_{d}$}}  & \multicolumn{4}{c}{\bf{Metrics}}\\
		\cline{2-5}	&Dice[\%] 	 	&Jaccard[\%]		&ASD[voxel] 		&95HD[voxel] \\ \hline
		DTML($\mathcal{L}_{seg}$/$\mathcal{L}_{dis}$)      & 89.72     & 81.55     & 1.97     & 7.64       \\
		DTML($\mathcal{L}_{seg}$/$\mathcal{L}_{mask}$)     & 90.12     & 82.14     & 1.82     & 7.01       \\
		DTML($\mathcal{L}_{seg}$/$\mathcal{L}_{dis}+\mathcal{L}_{mask}$)  & 89.89     & 81.82     & 1.93     & 6.61       \\ \hline \hline
	\end{tabular}
\end{table}

\begin{table}[t]
	\caption{Ablation analysis of our dual-task mutual learning framework. } \label{Table2}
	\centering
	\renewcommand\arraystretch{1.1}
	\begin{tabular}{c|c|c|c|c|c|c}
		\hline 	\hline
		\multirow{2}{*}{\bf{Method}} & \multicolumn{2}{c|}{\textbf{Scans used}} & \multicolumn{4}{c}{\bf{Metrics}}\\
		\cline{2-7}					&Labeled 	&Unlabeled	&Dice[\%] 	 	&Jaccard[\%]		&ASD[voxel] 		&95HD[voxel] \\ \hline
		$M_{s}$ only          & 16            & 0             & 86.03     & 76.06       & 3.51       & 14.26        \\
		$M_{d}$ only          & 16            & 0             & 88.69     & 79.91       & 3.12       & 11.62    \\ 
		DTML       & 16            & 0              & 89.17 & 80.68 & 2.13 & 7.82       \\ \hline \hline
	\end{tabular}
\end{table}

\subsection{Ablation Analysis} \label{ablation}

We conduct detailed experimental studies to examine the effectiveness of our proposed framework.
For supervised loss of $M_{d}$, $L_{2}$ loss between the output signed distance maps and transformed distance maps of ground truth named $\mathcal{L}_{dis}$, and segmentation loss between transformed segmentation mask of output distance maps and binary ground truth mask named $\mathcal{L}_{mask}$, can both be employed for the training. We make an comparison between different settings for $M_{d}$. It can be observed from Table \ref{Table1} that using $\mathcal{L}_{mask}$ for supervised loss can obtain higher performance compared with using $\mathcal{L}_{dis}$ or their sum. 
Experimental results demonstrate that direct supervision based on segmentation masks for both $M_{s}$ and $M_{d}$ is the best practice for our framework, since minor differences on signed distance maps may somehow mislead the training.

\begin{figure}[!t]
	\includegraphics[width=12.5cm]{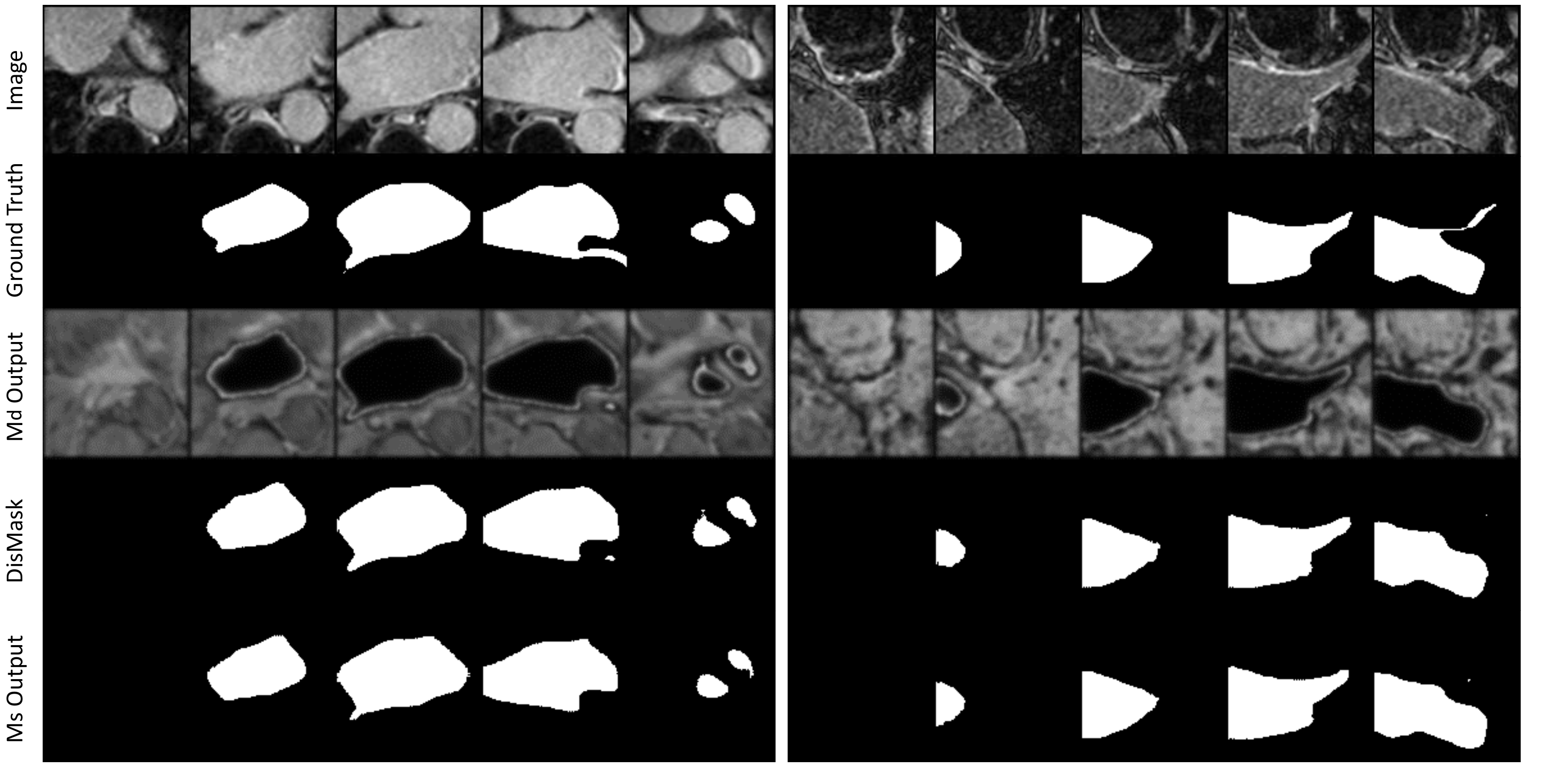}
	\caption{Visual comparison of output signed distance maps ($M_{d}$ Output), transformed segmentation masks (DisMask) and output segmentation maps ($M_{s}$ Output) in our framework. The first two rows are corresponding image and ground truth.}
	\label{Fig3}
\end{figure}

\begin{figure}[!t]
	\includegraphics[width=12cm]{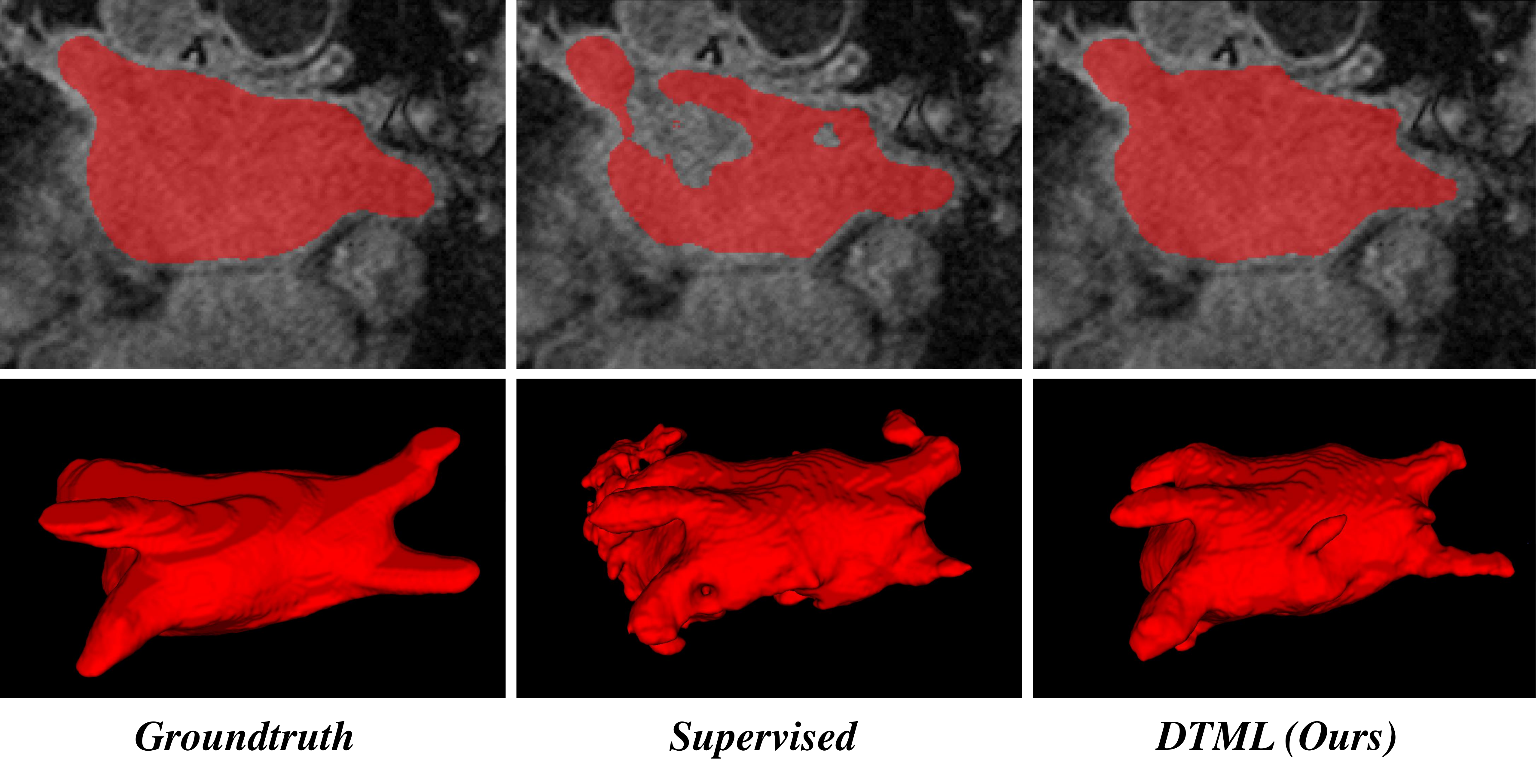}
	\caption{Examples of 2D and 3D visual comparison of different segmentation results.}
	\label{Fig2}
\end{figure}

Besides, to analyze the effectiveness of our method, we conduct experiments to implement our method with only labeled cases and remove the cross-task consistency loss between dual-task networks for comparison.
The first and second rows in Table \ref{Table2} are segmentation results of $M_{s}$ and $M_{d}$ training independently. We can observe that our method significantly outperforms both tasks training independently on all metrics. Paired T-test shows that the improvements are statistically significant at
$p < 0.05$ compared with both $M_{s}$ and $M_{d}$, validating the effectiveness of our mutual learning framework.
Figure \ref{Fig3} presents some examples of output from different network in our framework. It can be observed that both networks can achieve promising segmentation results. Besides, the transformed segmentation masks and output segmentation maps are slightly different due to the task difference, which enables the knowledge transfer between dual-task networks .
In Figure \ref{Fig2}, we show some segmentation examples of supervised method and our proposed semi-supervised method for visual comparison. We can observe that our segmentation results have higher overlap ratio with the ground truth.

\begin{table}[!t]
	\caption{Quantitative comparison between our method and other semi-supervised methods. All the models use the same V-Net as the backbone. The first and second rows are upper-bound performance and fully supervised baseline. Experimental results demonstrate that our method outperforms the state-of-the-art results consistently.} \label{Table3}
	\centering
	\renewcommand\arraystretch{1.1}
	\begin{tabular}{c|c|c|c|c|c|c}
		\hline 	\hline
		\multirow{2}{*}{\bf{Method}} & \multicolumn{2}{c|}{\textbf{Scans used}} & \multicolumn{4}{c}{\bf{Metrics}}\\
		\cline{2-7}					&Labeled 	&Unlabeled	&Dice[\%] 	 	&Jaccard[\%]		&ASD[voxel] 		&95HD[voxel] \\ \hline
		V-Net (upper bound)     & 80            & 0              & 91.14     & 83.82        & 1.52        & 5.75         \\
		V-Net (lower bound)     & 16            & 0              & 86.03     & 76.06        & 3.51        & 14.26        \\ \hline
		ASDNet\cite{nie2018asdnet}      & 16            & 64             & 87.90     & 78.85        & 2.08        & 9.24         \\
		TCSE\cite{li2020transformation} & 16            & 64             & 88.15     & 79.20        & 2.44        & 9.57         \\
		UA-MT\cite{yu2019uncertainty}   & 16            & 64             & 88.88     & 80.21        & 2.26        & 7.32         \\
		DTC\cite{luo2020semi}           & 16            & 64             & 89.42     & 80.89        & 2.10        & 7.32         \\
		SASS\cite{li2020shape}          & 16            & 64             & 89.54     & 81.24        & 2.20        & 8.24         \\
		DoubleUnc\cite{wang2020double}  & 16            & 64             & 89.65     & 81.35        & 2.03        & 7.04         \\ \hline
		\textbf{DTML (Ours)}            & 16            & 64             & \textbf{90.12} & \textbf{82.14} & \textbf{1.82} & \textbf{7.01} \\ \hline \hline
	\end{tabular}
\end{table}

\subsection{Quantitative Evaluation and Comparison}

To demonstrate the effectiveness of our method, a comprehensive comparison with existing methods is conducted.
We evaluate our method in with comparisons to state-of-the-art semi-supervised segmentation methods, including ASDNet \cite{nie2018asdnet}, TCSE \cite{li2020transformation}, UA-MT \cite{yu2019uncertainty}, DTC \cite{luo2020semi}, SASS \cite{li2020shape} and Double-Uncertainty \cite{wang2020double}.
To ensure a fair comparison, we used the same V-Net backbone in these methods. 
As a contrast, we conduce experiments of V-Net under fully-supervised settings with 20\% and all labeled data as the lower-bound and upper-bound performances for the task. Compared with semi-supervised learning settings, only labeled scans are used for the lower-bound subtask and both labeled and unlabeled scans with annotations are used for the upper-bound subtask. 
The results of comparison experiments are shown in Table \ref{Table3}.
As can be observed, by exploiting unlabeled data for training, our proposed method can produce significant performance gains and ontain obtain comparable results (90.12\% vs. 91.14\% of Dice) with the upper-bound performance.
In addition, our method achieves better performance on all the evaluation metrics compared with state-of-the-art segmentation methods.


\section{Conclusion}

In this paper, we propose a novel dual-task mutual learning framework for semi-supervised medical image segmentation. Our method can effectively leverage abundant unlabeled data by encouraging the output consistency of two tasks: learning region-based shape constraint and learning boundary-based surface mismatch.
By jointly learning the semantic segmentation and signed distance maps of targets and building task-level regularization, our framework can enforce the geometric shape constraint and collaboratively learn more reliable information from unlabeled images throughout the training procedure. Comprehensive experimental analysis demonstrates the effectiveness of our proposed method and significant improvement compared with state-of-the-art semi-supervised segmentation methods.

\section*{Acknowledgment}

This work is supported by the National Key Research and Development Program of China (2016YFF0201002), the University Synergy Innovation Program of Anhui Province (GXXT-2019-044), and the National Natural Science Foundation of China (61301005).

%
%
%
\bibliographystyle{splncs04}
\bibliography{reference}

\end{document}